\documentclass[twoside,10pt,letterpaper]{newFNLstyle}
\usepackage{graphics}
\usepackage{cite}

\newcommand{\bm }{ {\overline{ m}} }
\newcommand{\M}{{\cal M}}
\newcommand{\rhoL}{{\rho_\Lambda}}

\newcommand{\bx}{{y}}
\renewcommand{\d}{{\rm d}}

\newcommand{\bi}{{\rm bi}}
\newcommand{\half}{{\frac{1}{2}}}

\newcommand{\BEQ}{\begin{eqnarray}}   
\newcommand{\EEQ}{\end{eqnarray}}   
\newcommand{\BEA}{\begin{eqnarray}}   
\newcommand{\EEA}{\end{eqnarray}}   
\newcommand{\nn}{\nonumber }   
\renewcommand{\d}{{\rm d}}

\begin{document}

\volnumpagesyear{0}{0}{000--000}{2008}
\dates{received date}{revised date}{accepted date}

\title{EXACT SOLUTION FOR THE INTERIOR OF A BLACK HOLE}

\authorsone{THEO M. NIEUWENHUIZEN}

\affiliationone{Institute for Theoretical Physics, University of Amsterdam}

\mailingone{Valckenierstraat 65, 1018 XE Amsterdam, The Netherlands}

\maketitle

\markboth{Exact solution for the interior of a black hole}{Nieuwenhuizen}

\pagestyle{myheadings}


\begin{abstract}
Within the Relativistic Theory of Gravitation it is shown that  
the equation of state $p=\rho$ holds near the center of a black hole.
For the stiff equation of state $p=\rho-\rho_c$ the interior metric is 
solved exactly. It is matched with the Schwarzschild metric, which 
is deformed in a narrow range beyond the horizon.
The solution is regular everywhere, with a specific shape at the origin.
The gravitational redshift at the horizon remains finite but is large, 
$z\sim 10^{23}$$M_\odot/M$. Time keeps its standard role also in the interior.
The energy of the Schwarzschild metric, shown to be minus infinity in 
the General Theory of Relativity, is regularized in this setup, 
resulting in $E=Mc^2$.
\end{abstract}

\maketitle

\section{Introduction}

Black holes (BH's) have fascinated mankind because of their predicted properties:
at the classical level not even light can escape; 
the mass is located in the center;
the role of time in the interior 
is played by space and vice versa; on the quantum level they radiate as a thermal 
body via the Hawking mechanism.

The structure of galaxies  is related to astrophysical BH's,
galaxies are believed to have a supermassive central 
black hole with mass equal to $0.12\%$ of the bulge mass.~\cite{BHmassBulgeMass}
The BH in the center of our own Galaxy has ``only" four million solar masses.
Some believe that our Universe is actually the inside of a giant black hole.
~\cite{BHUniverse}  

As the above mentioned theoretical aspects are physically difficult to understand, 
an important question is: Are they perhaps unrealistic, and should the
behavior of physical BH's not be sought in a modified version of the General 
Theory of Relativity (GTR)? One such candidate is the Relativistic Theory of Gravitation
(RTG) proposed by Logunov.~\cite{LogunovBook,LogInflaton}
Following idea's of Rosen ~\cite{Rosen} and others, e.g. ~\cite{Deser,GPP},
it writes the Hilbert-Einstein Lagrangian 
in terms of a field in Minkowski space-time, extends it with 
the cosmological term $-\rho_\Lambda=-8\pi G\Lambda/c^2$ and with a bimetric coupling
$\half\rho_\bi\gamma_{\mu\nu} g^{\mu\nu}$
between the Minkowski metric $\gamma$ and the Riemann metric $g$.
~\cite{LogunovBook,NEPL}
The bimetric term breaks the general coordinate invariance of GTR, and allows only the 
harmonic gauge for the metric or the Lorentz gauge for the field. 
Since it will be cosmologically small, 
~\cite{LogInflaton} RTG will have the same
content as GTR for all standard observable effects in the solar system and galactic
problems. Differences may arise near singularities, in particular, for black holes 
and in cosmology.

An important question is the value and, in particular, the sign of the new parameter 
$\rho_\bi$. Logunov pointed out that the choice $\rho_\bi=\rho_\Lambda$ cancels the
zero point energies of both terms, allowing to have far away from matter
just a Minkowski space. 
When RTG was formulated, good data for $\rho_\Lambda$ were not available, so it was
natural to choose $\rho_\bi<0$, with a graviton mass $m_g=\sqrt{-\rho_\bi/16\pi}$. 
To make up for the 
observed positive value of the cosmological constant, an inflaton field can be
added.~\cite{LogInflaton}
In ~\cite{NEPL} we have considered the situation where $\rho_\bi=\rho_\Lambda>0$,
its value being set by the present cosmological data, making an inflaton field obsolete. 
This leads to a tachyonic graviton, but
its tachyonic nature sets in only at the Hubble scale, where not individual 
gravitons but the whole Universe is relevant. 

The bimetric coupling regularizes the infinite redshift at the horizon of black holes
~\cite{LogunovBook,LogInflaton}. In a previous approach we presented a scaling behavior 
near the horizon: coming from the outside, the time-time and radial-radial 
components of the metric tensor follow the Schwarzschild shape, 
but they cross over to an exponential decay in the interior.\cite{NEPL} 
They do not change sign, and thus leave for time
its standard role, which is physically appealing: BH's are then extreme objects, 
but still behave similar to normal ones.  
Indeed, with the Killing vector remaining time-like in the interior, 
the Hawking mechanism does not work, as it neither does for Newton stars.

In \cite{NEPL} the solution in the interior was considered on the basis of
present consensus: all matter in or very near the origin. 
We modeled this by a very-low-pressure equation of state.
In our follow up studies, we have 
realized that this approach is inconsistent.
Here we report about the opposite case: matter spreads throughout the BH. 
It is modeled by the stiff equation of state
$p=\rho-\rho_c$. In GTR this shape is known to have simplifying features ~\cite{HESpaper}
and this appears to carry over to RTG.

\section*{2. Setup of the problem}

Static spherically symmetric bodies have a metric 
\BEQ \label{sphersymmetric}
\d s^2=U(r)c^2\d t^2-V(r)\d r^2-W^2(r)(\d\theta^2+\sin^2\!\theta\d\phi^2\!).\EEQ 
and lead in RTG to the $00$ and $11$ Einstein equations
\BEQ\label{Ein00} 
&&\frac{1}{W^2}-\frac{W'{}^2}{VW^2} -\frac{2W''}{VW}
+\frac{V'W'}{V^2W}=\frac{8\pi G}{c^4}
\rho_{\rm tot},
\nn \\ 
\label{Ein11} 
&& -\frac{1}{W^2}+\frac{W'{}^2}{VW^2} +\frac{U'W'}{UVW}
=\frac{8\pi G}{c^4}p_{\rm tot},
\EEQ
where the total density and pressure have the form
\BEQ \label{rhotot=}
\rho_{\rm tot}&=&\rho+\rhoL
+\frac{\rho_\bi}{2U}-\frac{\rho_\bi}{2V}-\frac{\rho_\bi r^2}{W^2},\nn\\
p_{\rm tot}&=&
p-\rhoL+\frac{\rho_\bi}{2U}-\frac{\rho_\bi}{2V}+\frac{\rho_\bi r^2}{W^2}.
\EEQ
The harmonic constraint reads
\BEQ \label{harm}
\frac{U'}{U}-\frac{V'}{V}+4\frac{W'}{W}=\frac{4rV}{W^2}.
\EEQ
Taking $G=c=1$ we define the mass function $\M(r)$ by 
\BEQ \label{VMdef} V=\frac{W'{}^2}{1-2\M/W}.\EEQ
The Einstein equations can now be written as
\BEQ \label{M'=}
\M'=4\pi W' W^2\rho_{\rm tot},\quad
\frac{W-2\M}{2UW^2}\frac{U'}{ W'}-\frac{\M}{W^3}=
4\pi \,p_{\rm tot}.
\EEQ

We shall neglect $\rho_\Lambda$ and, because $U\ll{\rm min} (V,1)$,
only keep the $\rho_\bi/U$ terms. Indeed, they are responsible for the
scaling behavior near the horizon.~\cite{LogunovBook,NEPL} 
The fundamental assumption of the present Letter is that they also 
determine the shape in the interior of the BH.

To start, let us neglect matter and suppose that $U\sim W^2$. 
Eq. (\ref{M'=}) first brings $\M\sim W$ and then sets $U=8\pi\rho_\bi W^2$. 
The energy conservation condition reads
\BEQ (\rho+p)U'+2p'U=0.
\EEQ
If we assume that $p$ and $\rho$ are bounded at small $W$, 
then the result $p'\sim- W'/W$ leads to a logarithmic divergence,
 in conflict with the assumed boundedness. 
The case $p=\kappa\rho$ implies $\rho\sim W^{-1-1/\kappa}$. 
For every $\kappa$ ($0\le\kappa\le1$) this is more singular than 
the presumed leading term $\rho_\bi/(2U)\sim W^{-2}$. 
Thus for a singular RTG solution
one cannot treat matter as a perturbation, rather  
one must take $p\approx \rho$ at large $\rho$, that is, near the origin.

Let us therefore consider the stiff equation of state~\cite{Zeldovich}
\BEQ p=\rho-\rho_c,\quad r<R;\qquad \rho=p=0,\quad r>R,\EEQ
From energy conservation there arises the shape
\BEQ \label{rhopU} 
p=\half \rho_c(\frac{U_c}{U}-1),\qquad
\rho=\half \rho_c(\frac{U_c}{U}+1),
\EEQ
where the subscript $c$ denotes the point $r=R$.

Notice that the vacuum equation of state, $p=-\rho$ is covered
in case they are homogeneous. It appears in the limit $U_c\to0$, $x_c\to0$.

\section*{3. Exact solution for the interior}

The problem now leads to a similar exact solution. In terms of 
\BEQ \label{x=}
x= \frac{W}{W_1},\qquad 
W_1=\sqrt{\frac{3}{8\pi\rho_c}},\EEQ
the result $U=8\pi(\rho_cU_c+\rho_\bi)W^2$ may be written as 
\BEQ \label{xc}
U=U_c\frac{x^2}{x_c^2},\qquad 
x_c=\sqrt{\frac{\rho_cU_c}{3(\rho_cU_c+\rho_\bi)}}.
\EEQ
while density and pressure read
\BEQ \rho=\half\rho_c(\frac{x_c^2}{x^2}+1),\qquad
p=\half\rho_c(\frac{x_c^2}{x^2}-1).\EEQ
We may define at the point $x=1$ 
\BEQ \kappa_1=\frac{p(x=1)}{\rho(x=1)}=\frac{x_c^2-1}{x_c^2+1}.
\EEQ
This allows to express
\BEQ \label{rhobisign}
\rho_cU_c=\frac{3x_c^2\rho_\bi}{\rho_c(1-3x_c^2)}
=-\frac{3(1+\kappa_1)}{2(1+2\kappa_1)}\rho_\bi.
\EEQ
Both factors in the left hand side being positive
($\rho_c$ because $p\le\rho$; $U_c$ to avoid a horizon $U=0$), 
it is seen that the sign of $\rho_\bi$ is set by the physically allowed value of $\kappa_1$. 
For ``classical'' matter, it is natural to assume that negative pressures do 
not occur, and that $p=0$ at the horizon, so $\kappa_1=0$. 
This is the case to be considered from now on. 
(Quantum matter may have $\kappa_1$ down to $-1$; We come back to this in the discussion.) 
The bimetric and cosmological coupling constants are the assumed to be negative.
We now have
\BEQ \label{MxVx}
\M=\frac{W_1}{4}x(1+x^2),\qquad V=\frac{2W_1^2x'{}^2}{1-x^2}.
\EEQ
The harmonic constraint thus brings
\BEQ 
2\frac{x'}{x}-2\frac{x''}{x'}-\frac{2xx'}{1-x^2}
+4\frac{x'}{x}=\frac{8rx'{}^2}{x^2(1-x^2)}.
\EEQ
Going to the inverse function $r(x)$ makes it linear,
\BEQ x^2(1-x^2)r''+x(3-4x^2)r'=4r.
\EEQ
Let us define the conjugate variable
\BEQ\label{y=} 
\bx=\sqrt{1-x^2},
\EEQ
The solution is then remarkably simple,

\BEQ\label{rx=}
r=r_1(1+\frac{y}{\sqrt{5}}) x^{\sqrt{5}-1}(1+y)^{-\sqrt{5}},
\EEQ
where $r_1$ is the value at $x=1$.  Near that point one has
$r=r_1(1-\frac{4}{\sqrt{5}}\bx)$. 
We can now derive from (\ref{MxVx}) and (\ref{rx=}),
\BEQ
W'&=&\frac{W_1\sqrt{5}}{4r_1}x^{2-\sqrt{5}}y(1+y)^{\sqrt{5}},\nn\\
\qquad \label{Vx=}
V&=&
\frac{5W_1^2}{8r_1^2}x^{4-2\sqrt{5}}(1+y)^{2\sqrt{5}}.
\EEQ

Our solution is thus completely explicit.
At the origin it exhibits the singularities known
for the stiff equation of state,\cite{HESpaper,LogInflaton}
\BEQ U=\bar U_1r^{\gamma_\mu},\quad
V=\half\gamma_\mu^2\bar W_1^2r^{\gamma_\mu-2},\quad
W=\bar W_1r^{\half\gamma_\mu},\EEQ
where $\gamma_\mu=\half(\sqrt{5}+1)$ is the golden mean.
But if we take $W$ as the coordinate, we have the Riemann metric
\BEQ
\d s^2=\frac{U_c}{W_1^2x_c^2}W^2\d t^2-\frac{2\d W^2}{1-{W^2}/{4M^2}}-
W^2\d\Omega^2,
\EEQ
in the interior of the BH.
It clearly is regular at its origin, with the factor $2$ coding the above singularities,
and a coordinate infinity (but not a change in signs) at the horizon.

\section*{4. Matching with the exterior}

Well away from matter, the harmonic constraint brings for $W$ the
Schwarzschild shape $W_S=r+M$, 
where $M\equiv\M(R)$ is the mass, basically equal to the mass as observed 
at infinity. This implies via Eq. (\ref{VMdef})
\BEQ V_S=\frac{1}{1-2M/W_S}=\frac{r+M}{r-M}=\frac{1}{U_S}.
\EEQ

In GTR  one has $\rho_\bi=0$, so Eq. (\ref{xc}) gives $x_c=1/\sqrt{3}$.
Taking $W$ as coordinate sets $W'\to1$ in (\ref{VMdef}), yielding
$V_c=3$ $=1/U_c$, $W_c=3M$. This solution is seen as a limit model 
for neutron stars.~\cite{HESpaper} 
But let us consider the matching problem in the present setup at some $r=R$, $x=X$.
 Since we have to match $W'=1$, it follows that $V=2/(1-X^2)$, 
implying $X^2=\half(R+3M)/(R+M)$.
On the other hand, equating $\M=\frac{1}{4}(1+X^2)W_S$ to $M$, yields as only
solution $R=M$, $X=1$, exceeding the presumed maximum $1/\sqrt{3}$.  
Thus, in the harmonic gauge the set
$(V,W,W')$ {\it cannot match the Schwarzschild values},
so this limit model of GTR must be distrusted. 

In general, we consider as {\it regular} any solution for which $\M(R)<\half W(R)$.
We  define a {\it black hole} as a solution 
for which $\M(R)\approx\half W(R)$.
For  a black hole Eq. (\ref{MxVx}) sets $x_c=1$ and 
Eq. (\ref{xc}) shows that this is possible within RTG, 
provided $\rho_\bi$ is negative,
which is the Logunov situation with a massive graviton.
Together with Eqs. (\ref{x=}), (\ref{rx=}) this amounts to 
\BEQ 
U_c 
=-\frac{3\rho_\bi}{2\rho_c},\quad W_1=2M,\quad 
\rho_c= \frac{3}{32\pi M^2},
\quad r_1=R.\,
\EEQ
Contrary the Schwarzschild philosophy, we demand that $V$ remains finite, 
as it occurs in (\ref{Vexact}) at $\bx=0$. Eq. (\ref{VMdef}) then offers 
an equivalent manner to characterize a BH,
\BEQ\label{BHcond}
 \textit{Criterion for black hole horizon:}\qquad W'(R)\approx 0. 
\EEQ

It is handy to introduce the inverse length $m_g$ (``graviton mass'') 
and the dimensionless small parameter $\bm_g$,
\BEQ \rho=-\frac{m_g^2}{16\pi},\qquad \bm_g=m_gM\le
1.5\, \,10^{-23}\frac{M}{M_\odot}, \EEQ
where we took the estimate from Ref.~\cite{LogInflaton}.
We shall therefore have the values at the horizon $R$
\BEQ \label{horizoncond}
U_c=\bm_g^2,\quad 
V_c=\frac{5}{2}\frac{M^2}{R^2},
\quad W_c=2M,\quad W'_c=0.
\EEQ
These values look worrying, as they are far from Schwarzschild's, 
even when $r$ is near $M$ (e.g., $W_S'=1$).
The problem nevertheless appears to be consistent.
Beyond $R$ we need the deformation of the Schwarzschild metric near the horizon,
caused by the bimetric coupling. It was studied 
by Logunov and coworkers~\cite{LogunovBook,LogInflaton}, and  
an elegant scaling form for small $\bm_g$ was presented by us, \cite{NEPL} 
\BEA\label{scaling}
r&=&M
\frac{1+\eta(e^\xi+\xi+r_0)}{1-\eta(e^\xi+\xi+r_0)},
\qquad
 U = \eta e^\xi, \\
V &=&\frac{e^ \xi}{\eta(1+e^\xi)^2},\qquad 
W=\frac{2M}{1- \eta (e^\xi+ w_0)+\bm_g^2\xi}.\nn
\EEA
Here $\xi$ is the running variable, $\eta$ a small scale 
and $r_0$ and $w_0$ parameters.
Coming from the outside, $ e^\xi= {\cal O}(1/\eta)$, the functions follow 
 the Schwarzschild shapes, with small corrections,
but they branch off for $e^\xi={\cal O}(1)$. 
 At $\xi=0$ the function $V$ has a maximum 
$1/(4\eta)$, while $U$ has already gone down to $\eta$.  
Going further to the inside, for $\xi\ll-1$, $U$ and $V$ both decay 
exponentially over a very short distance, $\delta r=2\eta M$.
It is this exponential decay that will provide the opportunity
to match the seemingly very different behaviors near the horizon.
Matching $U$, $V$ and $W$ with the boundary values (\ref{horizoncond}) of
the interior solution, we find that
\BEQ \label{zetasol=}
&&\eta =\sqrt{\frac{2}{5}}\,\bm_g, \qquad
e^\xi=\sqrt{\frac{5}{2}}\,\bm_g,\qquad \nn \\
&&w_0= \sqrt{\frac{5}{2}}\,\bm_g\,\left(\log\bm_g+\half\log{\frac{5}{2}}-1\right).
\EEQ
In this regime Eq. (\ref{scaling}) yields
\BEQ W'(r)=\frac{\eta e^\xi-\bm_g^2}{\eta(e^\xi+1)}. \EEQ
The values (\ref{zetasol=}) confirm  that $W'(R)=0$, at the considered order $\bm_g$.
This property nicely settles a subtlety. 
The scaling shape of $W'$ becomes negative below a certain $r$,~\cite{NEPL}  
a fact erroneously interpreted as self-repulsion.~\cite{LogunovBook,LogInflaton} 
If matter is taken into account, it induces
a $W'>0$ in the interior, which goes to zero at the horizon. 
This matches the zero coming from the outside.
All by all, one thus has $W'>0$, except for $W'(R)=0$, see Fig. 1.

To fix the parameters $r_0$ and $w_0$ of Eq. (\ref{scaling}), we need to 
see which effects they bring at finite $r$. Around the Schwarzschild solution 
(neglecting for now $\rho_\bi$ and $\rho_\Lambda$)
there are four perturbative modes. The first, 
\BEQ \label{dUVW1=}
\delta U_1&=& \frac{(rL-M)M}{(r+M)^2},\quad
\delta V_1= \frac{rL-M}{r-M}-\frac{LM^2}{(r-M)^2}, \nn\\
\delta W_1&=& \half(rL-M),\qquad L\equiv\half\log\frac{r+M}{r-M},
\EEQ
involves the logarithm of $V_S=1/U_S$.
The second mode, $(\delta U_2,\delta V_2,\delta W_2)=\partial(U_S,V_S,
\\ W_S)/\partial M$,
relates to a shift in the mass;
the third, ($U_S,0,0$), rescales $U_S$, while the fourth, ($0,V_S,\half W_S$),
rescales $V_S$ and $W^2_S$.
We require that the second is absent at order $\bm_g$ and,
to keep the proper behavior at infinity, that the
third and fourth are absent to all orders. At order $\bm_g\sim\eta$
this may be imposed by analyzing the behaviors of $U$, $V$ and $W$
of Eqs. (\ref{scaling}) for $r>M$, with $r-M$  small but fixed, yielding
\BEQ r_0=2+\log\eta,\qquad w_0={\cal O}(\bm_g),
\EEQ
the latter being in agreement with (\ref{zetasol=}).
The logarithmic mode (\ref{dUVW1=}) remains with prefactor $8\eta$, 
and it is coded in the terms linear in $\xi$ of  Eq. (\ref{scaling}).
It is a finite distance, nonperturbative effect of order $\bm_g\sim M\sqrt{-\rho_\bi}$.
Near the horizon, for $r-M\sim\bm_g M\ln(1/\bm_g)$, it signifies the 
onset of the deformation of the Schwarzschild metric.

\begin{figure}[htbp] 
\centering {\resizebox{5cm}{!}{\includegraphics{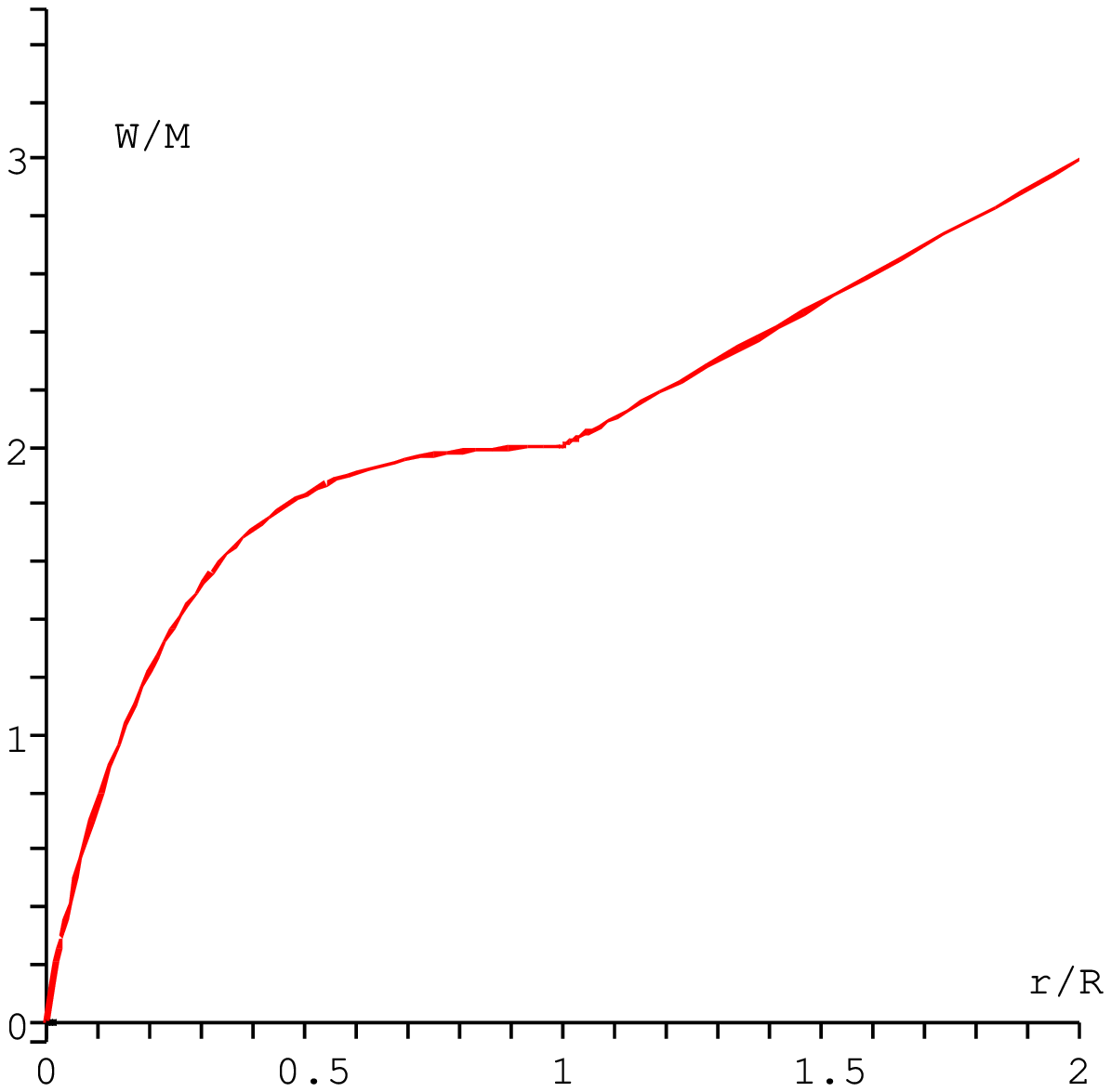}}}
\caption{ 
The metric function $W(r)$ has value $2M$ and slope zero at the horizon 
of a black hole.
$\bm=0.01$.}
\label{}
\end{figure}

\begin{figure}[htbp] 
\centering {\resizebox{5cm}{!}{\includegraphics{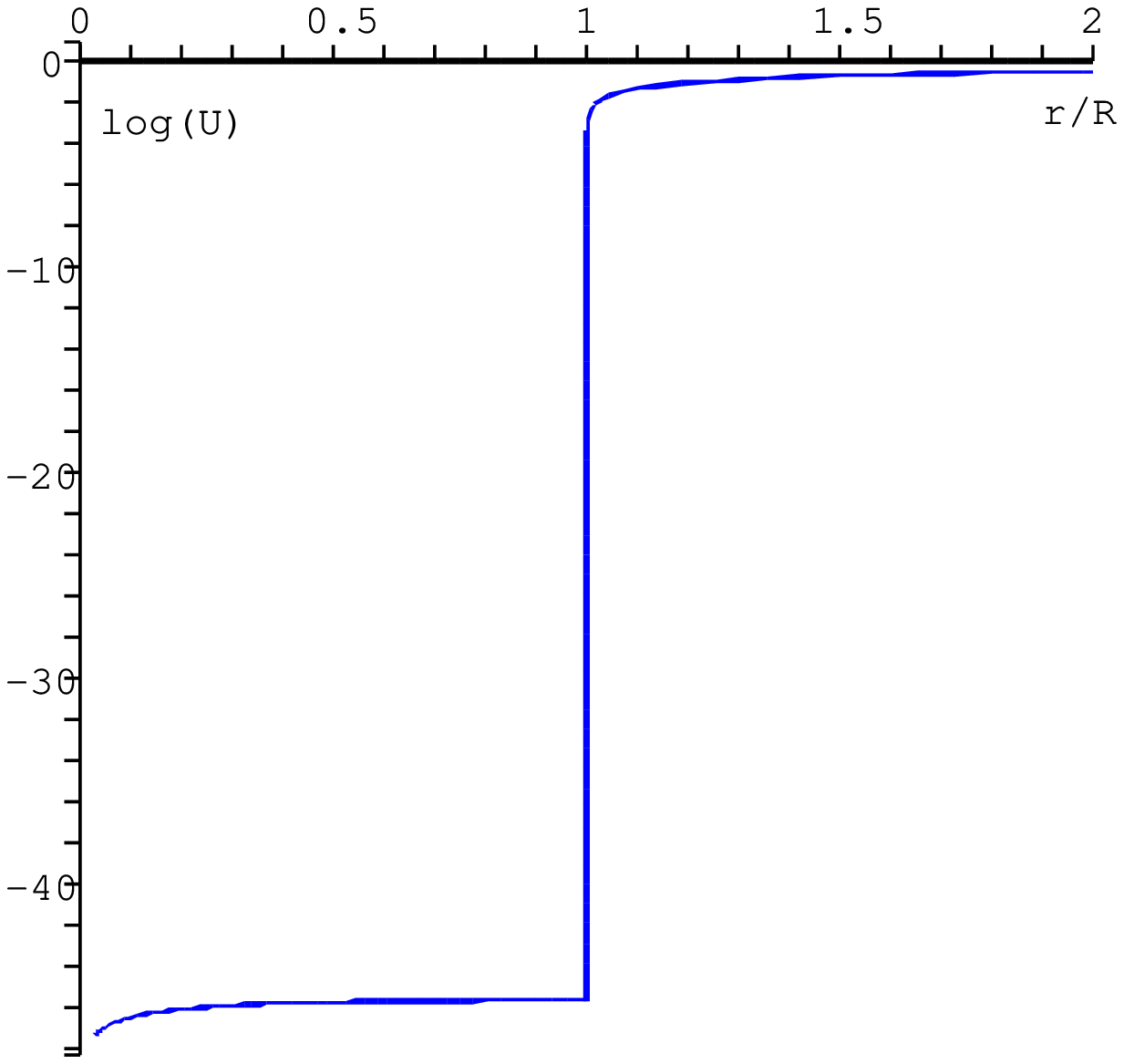}}}
\caption{ 
The metric function $U$ starts as a powerlaw
and becomes equal to $\bm_g^2$ at the horizon, beyond which it
grows exponentially towards the Schwarzschild shape.
$\bm_g=1.5\,\,10^{-23}$ corresponds to a
 one solar mass black hole.}
\label{figU}
\end{figure}

\begin{figure}[htbp] 
\centering {\resizebox{5cm}{!}{\includegraphics{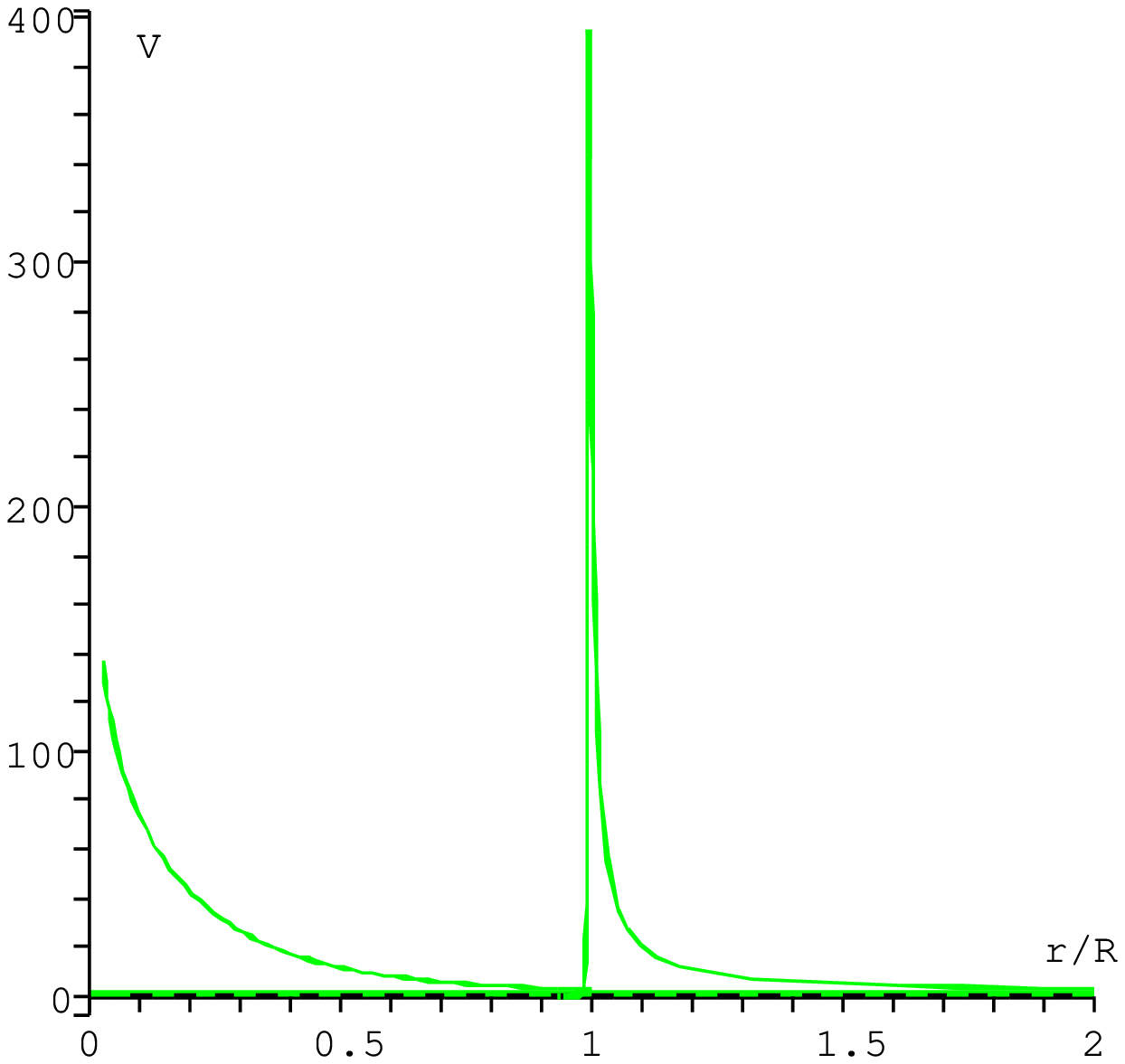}}}
\caption{ 
The metric function $V$ is of order unity in the interior, with a powerlaw divergence
at the origin. 
Beyond the horizon it grows exponentially towards a maximum 
of order $1/\bm_g$, 
after which it joins the Schwarzschild shape. 
 $\bm_g=10^{-3}$.}
\label{figV}
\end{figure}


From Eqs. (\ref{scaling}) and (\ref{zetasol=}) we 
get the horizon radius,
\BEQ \label{R=}
R=M\left[1+4\sqrt{\frac{2}{5}}\,\bm_g\,(\log\,\bm_g+1)\right]\,<\,M.
\EEQ

With the metric completely specified, we present plots of $U$, $V$ and $W$
in Figs. 1, 2, 3, respectively.

Having determined the leading scaling approach, we may 
go to next order in $\eta$ in the peak regime 
$e^\xi={\cal O}(1)$ and in the horizon regime where it is ${\cal O}(\bm_g)$.
This rather painful analysis will not be reported here.
We just mention the confirmation of our black hole condition 
$W'(R)=0$ at second order in $\bm_g$.

All by all, we may now rewrite the scaling form by eliminating $\xi$ in favor of $U$,
\BEA\label{scalingU}
r&=&M
\frac{1+U+\bm_g\sqrt{2/5}(\log U+2)}{1-U-\bm_g\sqrt{2/5}(\log U+2)},
\nn \\
V &=&\frac{U}{(U+\bm_g\sqrt{2/5})^2},\qquad  \\
W&=&\frac{2M}{1-U+\bm_g^2 +\bm_g^2\log(U/\bm_g^2)}.\nn
\EEA
This describes the free space region $r\ge R$, where $\bm_g^2\le U\le 1+{\cal O}(\bm_g)$.
At scale $r\sim1/m$ Newton's law is picks up the Yukawa factor $\exp(-mr)$,
due to the massive nature of gravitation in RTG.~\cite{LogunovBook}

The interior shape can also be expressed in $U$  as running variable, 
where it lies in the range $(0, \bm_g^2)$.
Due to Eqs. (\ref{xc}) and (\ref{y=}) it also holds that
\BEQ x=\frac{\sqrt{U}}{\bm_g},\qquad 
\bx=\sqrt{1-\frac{U}{\bm_g^2}}.
\EEQ 
The density and pressure read
\BEQ \label{rhopU} 
\rho=\half \rho_c(\frac{\bm_g^2}{U}+1), \qquad
p=\half \rho_c(\frac{\bm_g^2}{U}-1),
\EEQ
With $R$ given by Eq. (\ref{R=}), the locus is
\BEQ\label{rU} r=R(1+\frac{\bx}{\sqrt{5}}) 
(1-\bx)^{\half(\sqrt{5}-1)}(1+\bx)^{-\half(\sqrt{5}+1)},
\EEQ
and the other two metric functions read
\BEQ\label{Vexact} 
V=\frac{5M^2}{2R^2}(1-\bx)^{2-\sqrt{5}}(1+\bx)^{2+\sqrt{5}},
\quad W=2M\frac{\sqrt{U}}{\bm_g}.
\EEQ

The behaviors of the metric functions $U$, $V$ and $W$
are plotted in Figs. 1, 2 and 3, respectively.

\section*{5. Properties of the solution}
First of all, with $\rho$ and $p'$, the functions $U'$, $V'$, $W''$ 
are discontinuous at horizon. For $U'$ this is possible in 
the Eq. (2), because $W'(R)=0$.

The characteristic size of the deformation range of the Schwarzschild solution,
$\ell_{\rm deform}=r(\xi=1)-R$, is small,
\BEQ \ell_{\rm deform}\le
\sqrt{\frac{8}{5}}\,\bm_g\, M\,\ln\frac{e^{1+e}\sqrt{2}}{\bm_g\sqrt{5}}\approx 
1.6\, 10^{-18}\frac{M^2}{M_\odot^2}
\,{\rm m}.
\EEQ 
For one solar mass BH's this is comparable to
the Compton radius of the $W$ and $Z$ bosons, $\ell_W=2.45\,10^{-18}$m.

The gravitational energy density was discussed elsewhere.
\cite{BabakGrishchuk,NEPL}
For the metric (\ref{sphersymmetric}) it takes the form 
~\footnote{In static, spherically symmetric RTG the gravitational energy density is unique, 
``gravitational energy can be localized''.  This is implied by the harmonic constraint.
The residual gauge transformations allow only shapes that 
are unbounded at $0$ or $\infty$. As they would make the energy infinite,
they have to be discarded.
This holds both for regular solutions (stars) and for our BH solution.}

\BEQ \label{t00r=}
t^{00}&=&\frac{c^4W^2}{8\pi Gr^6}
\left(-\frac{r^2V'WW'}{V}+r^3V' -5r^2W'{}^2\right.\nn\\&+&
\left.\frac{2r^3VW'}{W}+8rWW'-2r^2V-3W^2\right).
\EEQ
At the origin it diverges as $r^{\sqrt{5}-5}$, which is
integrable.
\footnote{A different energy momentum tensor was proposed in Ref. ~\cite{GPP}.
Its energy density diverges at the origin as $r^{(\sqrt{5}-9)/2}$,
too singular to give a finite integral. For this reason we shall abandon it.}
The total energy density reads 
$\Theta^{00}=t^{00}+{VW^4}\rho_{\rm tot}/{r^4}$.
~\cite{NEPL} Its separate contributions are depicted in Fig. 2.

We can calculate the material energy.
A partial integration is needed to numerically tame the divergent 
behavior near $r=0$, that is exposed in Fig. 4. This brings 
\BEQ
U_{\rm mat}=4\pi \int_0^R\d rr^2\frac{VW^4}{r^4}\rho = 2228.830945\,M,
\EEQ
which is pretty large. Gravitational terms  are negative
and subtract $2227.830945\,M$ from this.
The bimetric term in $\rho_{\rm tot}$ contributes as
\BEQ
U_\bi=4\pi \int_0^R\d rr^2\frac{VW^4}{r^4}\frac{\rho_\bi}{2U}=
 -1372.93286\,M.
\EEQ
The gravitational energy inside the BH is
\BEQ
U_{\rm grav,\,int}=\int_0^R\d r 4\pi r^2t^{00}=-842.898079\,M.
\EEQ
Together they make up for $U_{\rm interior}=13 \,M$.

The gravitational energy density in the skin layer first has a large positive and then 
a large negative part, due to the term $r^3V'$, see Fig. 4 in the region around $r/R=1$. 
The integrated effect is obtained easily since the formulation of the Einstein equations 
in Minkowski space implies that the total energy density is a total derivative,
\BEQ \label{Theta=D}
\Theta^{00}=\frac{1}{4\pi r^2}\frac{\d}{\d r}
\left(\frac{VW^2}{2r}+\frac{W^4}{2r^3}-\frac{W^3W'}{r^2}\right).
\EEQ
where we have set $c=G=1$.
In the Schwarzschild regime\footnote
{For the Schwarzschild problem (matter-free Einstein equations in GTR),
in the standard gauge $W_S=r$, $U_S=1/V_S=1-2M/r$, 
we obtain $\Theta^{00}=-M^2/[2\pi r^2(r-2M)^2].$ 
Like (\ref{ThetaSS}), it has a quadratic divergence at the horizon, which is 
non-integrable and poses a so far overlooked problem for the Schwarzschild metric of GTR.}
so its  this combines into
\BEQ \label{ThetaSS}
\Theta^{00}=\frac{1}{4\pi r^2}\frac{\d}{\d r}\frac{M(r+M)^3(2r+M)}{2r^3(r-M)}.
\EEQ
From (\ref{Theta=D}), (\ref{ThetaSS}) and (\ref{horizoncond}), 
the region $R<r<\infty$ yields $U_{\rm exterior}=M(1-5-8+0)=-12 M.$
Together with the interior it makes up the total BH energy $U=Mc^2$.

\begin{figure}[htbp] 
\centering {\resizebox{8cm}{!}{\includegraphics{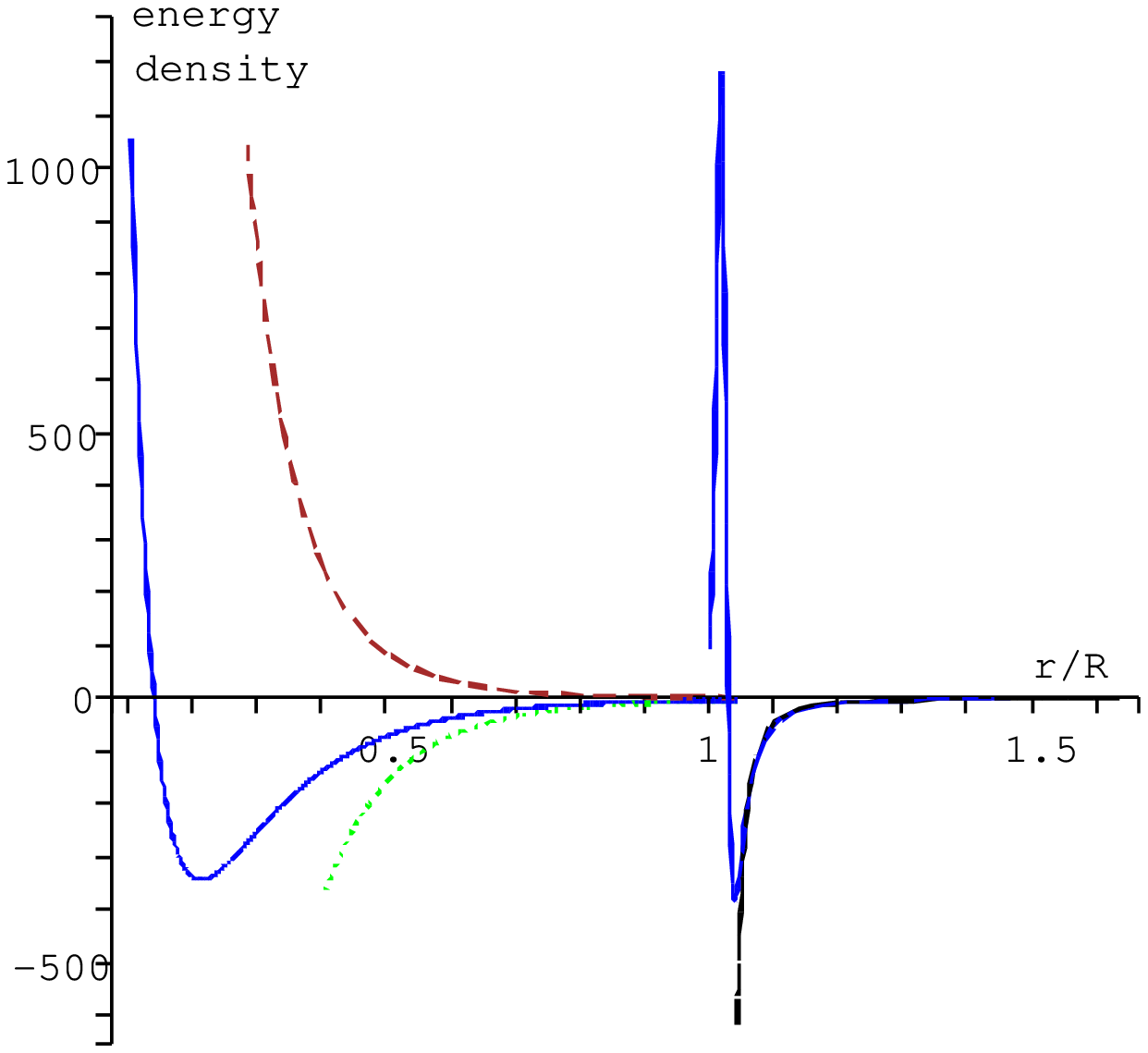}}}
\caption{
Full curve: The total energy density $\Theta^{00}$
as function of $r/R$ for  $\bm_g=0.00375$.
With $V'$, it is discontinuous at the horizon.
For realistic situations, very small $\bm_g$, the peak near the horizon
is much higher, narrower and  deeper on its right side.
Dashed: Material energy density $VW^4\rho/r^4$. 
Dots: Gravitational density $t^{00}+VW^4\rho_\bi/2Ur^4$ in the interior.
Dash-dots: Energy density of the Schwarzschild metric; in RTG
its divergence at the horizon is regularized by the full curve.}
\label{fig-1}
\end{figure}

\section*{6. Conclusion} 
We have considered a black hole in the Relativistic Theory of Gravitation. 
Previous findings that the bimetric coupling regulates the divergencies
of the Schwarzschild singularity, are extended to show that it sets
the behavior in the interior. Near the center, consistency requires that 
$p\approx\rho$ is unbounded.
For the case of the stiff equation of state
an exact and rather elegant solution is provided for the interior.
It matches the deformed Schwarzschild solution of the exterior.
For a BH of one solar mass, the deformation range is 
of the order of the Compton length of the $W$ and $Z$ bosons. 

Powerlaw singularities occur at the origin. 
This has been a reason to discard the problem,~\cite{LogInflaton}
but they disappear when the standard radial coordinate $W(r)$ is employed, 
rather than $r$ itself.  
Away from the origin, the solution is regular, 
and in particular also at the horizon, for any observer.
 The redshift at the horizon is
finite, though of the order $1/\bm_g\ge 10^{23}M_\odot/M$.
In the interior, time keeps it standard role. 
Hawking radiation is absent, and Bekenstein-Hawking entropy has no 
bearing. 


Open problems are to derive the radiation and to treat, for a given type of matter, 
the equation of state self-consistently with the metric, 
as was done here near the origin for any type of matter. 
Next, quantization of the field theoretic approach can be considered. 
One may also extend the approach to the Kerr-Newman black hole. 

Though we have elegantly described the interior of a black hole, 
we have not been able to settle definitively the question of the sign
of the bimetric coupling $\rho_\bi$. Indeed, while Eq. (\ref{rhobisign})
definitely leads to $\rho_\bi<0$ for non-negative pressures ($\kappa_1\ge 0$),
we cannot yet exclude the regime $-1\le\kappa_1 <-\half$, where a positive
$\rho_\bi$ would be required. Let us mention in this connection that the situation with 
the vacuum equation of state $p=-\rho$ ($\kappa_1=-1$) holding in the interior
was recently analyzed and stated to describe a ``gravastar'' in its Bose-Einstein
condensed ground state.~\cite{MM} 
We plan to investigate the connection between RTG black holes 
and Bose-Einstein condensation in the near future. In that situation
values of $\kappa_1$ near $-1$ may occur, in principle. Let us mention that
our exact metric remains valid for the vacuum equation of state,
that is, in the limit $\kappa_1\to-1$, $U_c\to0$, and $x_c\to 0$.

Having a complete solution at hand, we could also verify some general aspects
of the gravitational energy momentum tensor. We realized that the energy density
of the standard Schwarzschild metric of GTR has at the horizon a quadratic singularity.  
The related infinite gravitational energy is not apparent in the Riemann approach,
the singularity then being viewed as a coordinate singularity.~\cite{Petrov}
When making the step from GTR to RTG,
the divergence at the horizon gets regularized~\cite{NEPL},
and here we have seen that the total energy is finite, and equal to $Mc^2$, as it should.
We also verified that for static, spherically symmetric bodies,
residual gauge transformations within this sector would lead to infinite energies.
So they are forbidden, making the local energy density uniquely defined:
{\it energy and, in particular, gravitational energy, can be localized in RTG},
probably under more general conditions than reported here. 

The fact our black hole has no true horizon would justify as name: grey hole.
However, if observed black holes in the cosmos have a huge but finite redshift at their 
horizon, as described here, it is better to stick to the standard name. 
 
The resolution of the singular behavior at the horizon arises from the bimetric 
coupling, which acts as a mass-type term. This breaks the general coordinate invariance of GTR. 
Since $\rho_\bi$ is cosmologically small, this could normally play a role only in cosmology
-- but it still allows the $\Lambda$ Cold Dark Matter model.
Indeed, it brings no change of general relativistic effects in the solar system or for
gravitational radiation of binaries.
However, we have seen that the bimetric term does play a role at large redshifts, 
that is to say, near the horizon of Schwarzschild black holes and inside it.
This resolution of a singularity may be more general.

Returning to the black hole problem:
It is sometimes argued that our Universe may actually be the inside of a 
giant black hole.~\cite{BHUniverse} For that application, the Schwarzschild metric 
with all its matter in the center is not realistic, and
our setup with matter spread throughout the interior looks more natural.

\end{document}